\documentclass[symmetry,article,accept,pdftex,moreauthors]{Definitions/mdpi} 

\usepackage{latexsym}
\usepackage{amsfonts}
\usepackage{amsmath}
\usepackage{amssymb}

\newlength{\myboxsize}
\firstpage{1} 
\pubvolume{1}
\issuenum{1}
\articlenumber{0}
\pubyear{2025}
\copyrightyear{2025}
\externaleditor{Kazuharu Bamba} % More than 1 editor, please add `` and '' before the last editor name
\datereceived{29 January 2026} 
\daterevised{18 February 2026} % Comment out if no revised date
\dateaccepted{23 February 2026} 
\datepublished{ } 
\hreflink{https://doi.org/} % If needed use \linebreak
\Title{Choice of Quantum Vacuum for Inflation Observables}

\Author{Melo 
 Wood-Saanaoui $^{1}$, Rudnei O. Ramos $^{2,}$*\orcidB{} and Arjun Berera $^{1}$\orcidC{}}

\AuthorNames{Melo Wood-Saanaoui, Rudnei O. Ramos and Arjun Berera}

\address{$^{1}$ \quad Higgs Centre for Theoretical Physics, School of Physics and Astronomy, University of Edinburgh,\linebreak Edinburgh EH9 3FD, UK; e.wood-saanaoui@sms.ed.ac.uk (M.W.-S.); ab@ed.ac.uk (A.B.)\\
$^{2}$ \quad Departamento de Fisica Teorica, Universidade do Estado do
  Rio de Janeiro, 20550-013 Rio de Janeiro, RJ,
 Brazil}

\corres{Correspondence: rudnei@uerj.br}

\abstract{We investigate the modifications to inflationary observables that 
arise when adopting an $\alpha$-vacuum instead of the standard Bunch--Davies 
vacuum for quantum fluctuations during inflation. 
Within the Starobinsky inflationary model, we compute and compare the scalar 
spectral index, its running, and the running of the running arising from 
different choices of the initial vacuum state.
We further examine the energy scales associated with $\alpha$-vacua and argue 
that, for any number of extra spatial dimensions, the relevant scale can be 
truncated at the Hubble scale, $\sim$$\mathcal{O}(10^{13})\,\mathrm{GeV}$, 
without conflict with current Cavendish-type experimental bounds on 
sub-millimeter gravity (
$\sim$$250\,\upmu\mathrm{m}$). 
Our analysis demonstrates that the $\alpha$-vacuum is subject to stringent 
constraints as a viable de~Sitter-invariant alternative to the Euclidean 
(Bunch--Davies) vacuum, with the corrections that it induces in the inflationary 
observables being strongly limited by the latest Planck data.}

\begin{document}

\section{Introduction}

Inflation generically drives the universe into a quantum vacuum state 
in which quantum fluctuations seed the primordial curvature perturbations 
responsible for the anisotropies in the cosmic microwave background (CMB) 
and the formation of large-scale structures. The expression for the 
primordial power spectrum---from which the key inflationary observables 
are derived---depends sensitively on the choice of the quantum vacuum 
used to define the initial conditions for the perturbations. The conventional 
choice is the Bunch--Davies (BD) vacuum, defined by the condition
\begin{equation}
    \hat{a}_{k}\,|0\rangle = 0 \qquad \text{as} \qquad \tau \to -\infty \;,
\end{equation}
where $\tau$ is the conformal time ($d \tau = dt/a(t)$, with $a(t)$ the 
scale factor) and the mode functions asymptotically approach those of 
Minkowski spacetime in the far past~\cite{Kundu:2011sg}. This assumption---that 
the short-distance behavior of quantum fields in curved spacetime reduces to
Minkowski space in the infinite past---motivates the BD vacuum as the ``natural'' choice for inflation.

However, it has long been recognized that the BD state is not the unique 
de~Sitter-invariant vacuum. The seminal works of Mottola, Allen, and 
others~\cite{Allen:1985ux,Mottola:1983iq,Chernikov:1968zm} demonstrated 
that a one-parameter family of de~Sitter-invariant vacua exists, the 
so-called $\alpha$-vacua. These states are related to the Euclidean 
modes by a Mottola--Allen (Bogoliubov) transformation and reduce to the 
BD vacuum in the limit $\tau \to -\infty$. In the $\alpha$-vacuum, the 
transformation induces a natural momentum-space truncation at an energy scale 
$\Lambda$, above which the adiabatic notion of particles and the geometric 
interpretation of spacetime cease to be reliable. This reflects the 
expectation that trans-Planckian or stringy physics becomes important 
at sufficiently high energies, preventing the extrapolation of quantum 
field theory in curved spacetime to arbitrarily short wavelengths~\cite{Chen:2024ckx,Linde:1990flp}.

In this work, we do not interpret the $\alpha$-vacuum as a fundamental
alternative to the BD state but rather as an effective
parameterization of unknown ultraviolet (UV) physics.
Such modifications may arise from trans-Planckian effects, finite-time
boundary conditions, or pre-inflationary dynamics that prevent the
extrapolation of quantum field theory to arbitrarily short wavelengths.
In this sense, the $\alpha$-vacuum provides a controlled and
model-independent framework to quantify possible deviations from the
standard BD predictions.

In this work, we also focus on the late (slow-roll) stage of inflation, 
during which observable cosmological perturbations are generated and exit 
the Hubble radius. Our analysis is therefore insensitive to the details 
of the onset of inflation or possible pre-inflationary phases.

The modified initial conditions of the $\alpha$-vacuum feed directly 
into corrections to the primordial power spectrum. Consequently, inflationary 
observables such as the scalar spectral index $n_{s}$, its running 
$\alpha_{s} \equiv dn_{s}/d\ln k$, and the running of the running 
$\beta_{s} \equiv d^{2}n_{s}/d(\ln k)^{2}$ acquire contributions 
dependent on $\Lambda$. As expected, in the limit of a sufficiently 
high momentum cutoff, $\Lambda \to \infty$, or, more physically, 
$\Lambda \gg H$, these corrections vanish and the predictions 
continuously reduce to those of the standard BD vacuum.

A common assumption in effective field theory is to take the cutoff at the Planck scale,
\begin{equation}
    \Lambda \sim M_{\mathrm{Pl}} \sim 10^{18}~\mathrm{GeV} \;,
\end{equation}
which is motivated by the expectation that quantum-gravitational 
effects become dominant at this scale.  
Within this standard framework, trans-Planckian corrections are 
suppressed by powers of $M_{\mathrm{Pl}}^{-1}$ and therefore 
render any distinction between the BD and $\alpha$-vacua observationally 
negligible for CMB-relevant modes.

However, the Planck-scale choice for $\Lambda$ is not mandatory and should 
be regarded as an effective assumption rather than a fundamental requirement.  
In particular, scenarios with large extra dimensions provide a 
well-motivated setting in which the fundamental gravitational scale 
is lowered relative to $M_{\mathrm{Pl}}$, while remaining consistent 
with laboratory tests of gravity and cosmological observations~\cite{Anchordoqui:2023etp,Mohapatra:2000cm,Green:2002wk}.  
In such frameworks, the validity of the four-dimensional effective 
description naturally breaks down at scales below $M_{\mathrm{Pl}}$, 
allowing for a physically meaningful truncation at much lower energies. 
Accordingly, we consider the alternative possibility
\begin{equation}
    \Lambda \sim H_{\mathrm{infl}} \sim 10^{13}~\mathrm{GeV} \;,
\end{equation}
which corresponds to imposing the initial conditions during the slow-roll 
stage of inflation, where the quasi-de Sitter effective description of 
perturbations is valid (e.g., around Hubble radius crossing).  
This choice does not assume a specific UV completion but rather provides 
a conservative and self-consistent parameterization of unknown 
high-energy physics during inflation.  
As we show, even in this minimal scenario, the resulting $\alpha$-vacuum 
corrections are tightly constrained by current CMB data~\cite{Planck:2018vyg}, 
thereby placing strong bounds on deviations from the BD state.

The goal of this paper is to systematically quantify these corrections 
and determine to what extent $\alpha$-vacua constitute a viable alternative 
to the BD state. We evaluate the scalar spectral index, its running, and 
the running of the running in both vacua, with particular emphasis on 
their dependence on the cutoff scale $\Lambda$ and on the viability of low-scale truncations motivated by large-extra-dimension models. Throughout, we use the Starobinsky $R^{2}$ inflation model as an explicit case study, 
both for its phenomenological success and for its analytic tractability.

This paper is organized as follows. In Section~\ref{section2}, we review 
the $\alpha$-vacuum formalism and show how the scalar of curvature 
primordial power spectrum is modified.
In Section~\ref{section3}, we determine how the modified power spectrum 
causes corrections to the cosmological observables---in particular, to the spectral tilt and its runnings. We also discuss mechanisms for which the 
$\alpha$-vacuum correction might become relevant, e.g., in the case of 
extra dimensions. In Section~\ref{section4}, we present the numerical 
results of our analysis. {}Finally, in Section~\ref{conclusions}, we present 
our conclusions and final remarks about other possible applications of our results.

%%%%%%%%%%%%%%%%%%%%%%%%%%%%%%%%%%%%%%%%%%%%%%%%%%%%%%
\section{Formalism}
\label{section2}

In a quasi-de Sitter background, scalar perturbations are quantized by 
expanding the field operator in creation and annihilation operators 
associated with a chosen vacuum state. The conventional choice is the 
BD vacuum, defined by requiring that, deep inside the horizon, in the limit 
$\tau \to -\infty$, the spacetime appears Minkowskian and all modes reduce to positive-frequency plane waves \cite{Bunch:1978yq,Kundu:2011sg}:
\begin{equation}
    \hat{a}_{\vec{k}}\,|0_{\rm BD}\rangle = 0  \;.
\end{equation}
Note that the
 reference to a Minkowski spacetime in the far past should be understood 
in the effective sense that, for modes deep inside the Hubble radius, spacetime 
can be treated as locally flat.
Now, quantum field theory in curved spacetime does not generally admit 
a unique vacuum. In de Sitter space, the full $SO(4,1)$ symmetry allows 
an infinite family of invariant states, the so-called $\alpha$-vacua, 
first constructed by Mottola and Allen \cite{Mottola:1984ar,Allen:1985ux,Chernikov:1968zm}. 
These vacua arise from a Bogoliubov transformation of the BD mode functions $\phi^{\rm BD}_k(\tau)$:
\begin{equation}
    \phi_k^{(\alpha)}(\tau)
        = N_\alpha \left( e^\alpha\, \phi_k^{\rm BD}(\tau)
        + e^{\alpha^\ast} \phi_k^{\rm BD}(\tau)^\ast \right) \;,
\label{bogo}
\end{equation}
where $\alpha \in \mathbb{C}$ labels the vacuum and $N_\alpha$ ensures proper 
normalization. {}For $\alpha \to -\infty$, one recovers the BD state.

The BD vacuum is privileged because it is uniquely selected by requiring 
regularity and minimal energy for modes deep inside the Hubble radius, 
where spacetime can be treated as approximately Minkowskian. More general 
$\alpha$-vacua arise through Bogoliubov transformations, as in Equation~\eqref{bogo}, 
and can be interpreted as excited initial states relative to the BD vacuum. 
In this sense, transitions among different $\alpha$-vacua reflect changes 
in the initial conditions induced by unknown UV physics, while the BD 
vacuum represents the preferred infrared attractor.

\subsection{The Mottola--Allen Transformation}

The family of de Sitter-invariant $\alpha$-vacua may also be described 
through the Mottola--Allen transformation \cite{Allen:1985ux}, which 
expresses new mode functions $\tilde{\phi}_k(\tau)$ in terms of the Euclidean (BD) modes:
\begin{equation}
    \tilde{\phi}_k(\tau) = A\, \phi_k(\tau) + B\, \phi_k^\ast(\tau) \;,
\label{MAtransf}
\end{equation}
with \(A,B\) constants satisfying
\begin{equation}
    |A|^2 - |B|^2 = 1 \;,
\end{equation}
ensuring canonical commutation relations. The most general parameterization is
\begin{equation}
    A = e^{i\gamma}\cosh\alpha, 
    \qquad
    B = e^{i(\gamma+\beta)}\sinh\alpha \;,
\end{equation}
with $\alpha,\beta,\gamma \in \mathbb{R}$. A global phase is irrelevant, so one may 
set $\gamma = 0$, yielding
\begin{equation}
    \tilde{\phi}_k(\tau)
        = \cosh\alpha\, \phi_k(\tau)
        + e^{i\beta}\sinh\alpha\, \phi_k^\ast(\tau) ,
\end{equation}
with the BD vacuum corresponding to $\alpha = 0$, where $A = 1$ and $B = 0$. 
Any $\alpha \neq 0$ defines a different but equally $SO(4,1)$-invariant de Sitter state.

%%%%%%%%%%%%%%%%%%%%%%%%%%%%%%%%%%%%%
\subsection{Finite Initial Time, Momentum Cutoff, and the Form of \(e^\alpha\)}

Physical considerations motivate the imposing of the initial condition 
not at $\tau\to -\infty$ but at a finite time $\tau_0$, representing the 
earliest epoch where the effective field theory is valid. {}Following~\cite{Danielsson:2002kx,Xue:2008mk}, 
the Bogoliubov coefficient in Equation~\eqref{bogo} can be written as
\begin{equation}
    e^\alpha
        = -\, e^{-2ik\tau_{0}}
            \frac{i}{\frac{2\Lambda}{H} + i} \;,
\label{expalpha}
\end{equation}
where $\Lambda$ is the physical momentum cutoff. The modes satisfy $k = a(\tau_0)\Lambda$, 
and, using the relation $a(\tau_0) = -1/(H\tau_0)$, one obtains
\begin{equation}
    e^\alpha = -e^{-2i\Lambda/H}\,
        \frac{i}{\displaystyle \frac{2\Lambda}{H} + i } \;.
\end{equation}

\subsection{Bogoliubov Coefficients in Terms of the Cutoff Scale}

The alternative parameterization of the Bogoliubov coefficients used 
in~\cite{Danielsson:2002kx} is
\begin{equation}
    B_{k}
        = \gamma_{k}\, A_{k},
    \qquad
    \gamma_{k}
        = \frac{i}{2k\tau_{0} + i} \;.
\label{112}
\end{equation}

{}For $2k\tau_0 = -2\Lambda/H$, one then finds
\begin{equation}
\label{eq:ABgamma}
|\gamma_k|^2 = \frac{H^2}{4\Lambda^2 + H^2},
\qquad
|A_k|^2 = \frac{4\Lambda^2 + H^2}{4\Lambda^2},
\qquad
|B_k|^2 = \frac{H^2}{4\Lambda^2},
\end{equation}
which implies
\begin{equation}
\label{eq:ABsum}
|A_k|^2 + |B_k|^2 = 1 + \frac{H^2}{2\Lambda^2} \;.
\end{equation}

Combining Equation~\eqref{expalpha} with the modulus of \(\gamma_k\), one finds
\begin{equation}
    e^{\alpha+\alpha^\ast}
        = \frac{H^2}{4\Lambda^2 + H^2} \;.
\end{equation}
This identity allows the power spectrum to be expressed solely in terms 
of the physical cutoff \(\Lambda\).

We emphasize that our analysis is performed within the quasi-de Sitter 
approximation appropriate to slow-roll inflation. Although the full, rigorous 
characterization of admissible initial states in generic curved backgrounds is 
beyond the scope of this work, the adopted parameterization is widely used to 
capture leading deviations from the BD vacuum in a controlled and phenomenological manner. 

\subsection{Power Spectrum in the $\alpha$-Vacuum}

The spectrum of the curvature perturbation in terms of \(\gamma_k\) is \cite{Danielsson:2002kx,Giesel:2020bht}
\begin{equation}
    \mathcal{P}_{\mathcal{R}}(k)
        = \left( \frac{H^2}{2\pi\dot\phi} \right)^2
            \frac{
                1 + |\gamma_k|^2
                - \gamma_k e^{-2ik\tau_0}
                - \gamma_k^{\ast} e^{2ik\tau_0}
            }{
                1 - |\gamma_k|^2
            } \;.
\end{equation}

Using the relations
\begin{equation}
    \gamma_k^\ast e^{2ik\tau_0} = -e^\alpha,
    \qquad
    \gamma_k e^{-2ik\tau_0} = -e^{\alpha^\ast} \;,
\end{equation}
the spectrum becomes
\begin{equation}
    \mathcal{P}_{\mathcal{R}}(k)
        = \left( \frac{H^2}{2\pi\dot\phi} \right)^2
            \frac{
                1 + e^{\alpha+\alpha^\ast}
                + e^\alpha + e^{\alpha^\ast}
            }{
                1 - e^{\alpha+\alpha^\ast}
            } \;.
\end{equation}

\begin{figure}[H]
    %\centering
    \includegraphics[width=0.52\linewidth]{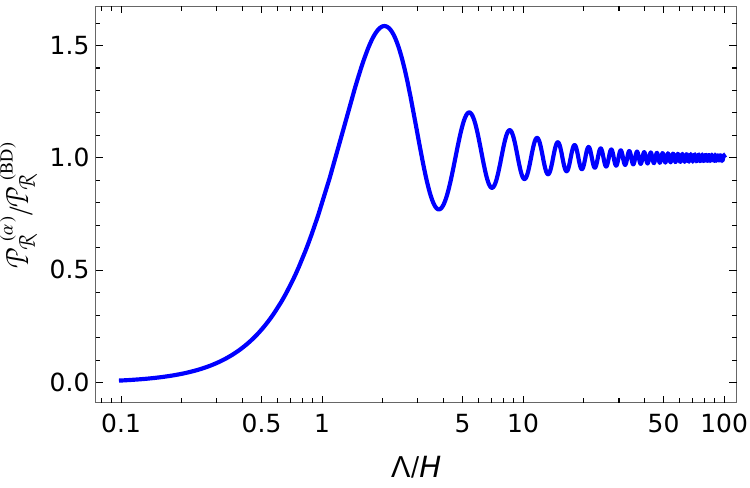}
    \caption{Comparison
 between the $\alpha$-vacuum prediction for the power 
    spectrum given by \mbox{Equation~(\ref{Pr})} and its BD limit, 
    $ {\cal P}_{\cal R}^{\rm (BD)}\equiv (H^4/(4 \pi^2 \dot \phi^2) $, 
    as a function of $\Lambda/H$. The BD result is recovered for $\Lambda\gg H$.}
    \label{fig0}
\end{figure}

{}Finally, substituting Equation~\eqref{expalpha}, we arrive at the explicit expression
\begin{equation}
    \mathcal{P}_{\mathcal{R}}(k)
        = \left( \frac{H^2}{2\pi\dot\phi} \right)^2
          \left[
            1
            + \frac{H^2}{2\Lambda^2}
            - \frac{H}{\Lambda}\,
                \sin\!\left(\frac{2\Lambda}{H}\right)
            - \frac{H^2}{2\Lambda^2}\,
                \cos\!\left(\frac{2\Lambda}{H}\right)
          \right] \;.
\label{Pr}
\end{equation}
This expression makes explicit the oscillatory cutoff-sensitive 
correction characteristic of the $\alpha$-vacua. In the limit $H/\Lambda \to 0 $, 
these corrections vanish and the BD result is recovered. In the next section, 
we examine how these modifications propagate into the scalar spectral index, 
its running, and the running of the running, and we assess observational 
constraints arising from trans-Planckian effects and scenarios with extra 
spatial dimensions. In {}Figure~\ref{fig0}, we illustrate how the $\alpha$-vacuum 
correction to the power spectrum in Equation~(\ref{Pr}) modifies the standard BD result. 

%%%%%%%%%%%%%%%%%%%%%%%%%%%%%%%%%%%%%%%%%%%%%%%%%%%%%%%%%%%%%%%
\section{Inflationary Observables in the \boldmath{$\alpha$}-Vacuum}
\label{section3}

In this section, we derive the corrected expressions for the scalar spectral index, 
its running, and the running of the running arising from the 
$\alpha$-vacuum-modified power spectrum in Equation~(\ref{Pr}).  
We then compare these expressions with both the standard Bunch--Davies predictions and with current observational constraints.

%%%%%%%%%%%%%%%%%%%%%%%%%%%%%%%%%%%%%%%%%%%%%%%%%%%%%%%%%%%%%%%
\subsection{The Scalar Spectral Index, Its Running, and the Running of the Running}

The scalar spectral index is defined in the usual way,
\begin{equation}
n_s(k)= \frac{d\ln\mathcal{P}_{\mathcal{R}}(k)}{d\ln k} \;,
\label{ns}
\end{equation}
with the running and running of the running given by
\begin{eqnarray}
\alpha_s &=& \frac{dn_s}{d\ln k} \;,
\label{alphas}
\\
\beta_s &=& \frac{d\alpha_s}{d\ln k} \;.
\label{betas}
\end{eqnarray}
These quantities are evaluated at the pivot scale and compared to observational datasets such as those from 
Planck and the Atacama Cosmology Telescope (ACT)~\cite{Planck:2018vyg,Planck:2018jri,ACT:2025fju,ACT:2025tim}.

{}For reference, in the standard $\Lambda$CDM model using Planck TT (TT, TE, EE) + lowE + lensing (68\% CL), we find~\cite{Planck:2018jri}
\begin{eqnarray}
  n_s &=& 0.9587\pm0.0056\quad (0.9625\pm0.0048) \;, \nonumber\\
  \alpha_s &=& 0.013\pm0.012\quad (0.002\pm0.010) \;, \nonumber\\
  \beta_s &=& 0.022\pm0.012\quad (0.010\pm0.013) \;.
\end{eqnarray}

At the Hubble radius crossing, we have $k_*=aH$, so, for $N=\ln a$, the number of $e$-folds
\begin{equation}
\frac{d\ln k}{dN}=1-\epsilon_H \;,
\label{dlnk}
\end{equation}
where $\epsilon_H=-\dot H/H^2$.  
In the slow-roll regime, we use $\epsilon_H\simeq\epsilon_V$, with the slow-roll 
coefficients defined in terms of the inflaton potential $V(\phi)$ 
as (note that we follow the conventions of Ref.~\cite{Planck:2013jfk})
\begin{eqnarray}
\epsilon_V &=& \frac{M_{\rm Pl}^2}{2}\!\left(\frac{V_{,\phi}}{V}\right)^2 \;,
\label{epsilonV}
\\
\eta_V &=& M_{\rm Pl}^2\frac{V_{,\phi\phi}}{V} \;,
\label{etaV}
\\
\xi_V^2 &=& M_{\rm Pl}^4\frac{V_{,\phi}\,V_{,\phi\phi\phi}}{V^2} \;,
\label{xi2V}
\\
\omega_V^3 &=& M_{\rm Pl}^6\frac{V_{,\phi}^2\,V_{,\phi\phi\phi\phi}}{V^3} \;,
\label{omega3V}
\end{eqnarray}
where $M_{\rm Pl} = (8\pi G)^{-1/2} \simeq 2.4\times10^{18}\,$GeV is the reduced Planck mass.

Applying Equations~(\ref{ns})–(\ref{betas}) to the modified power spectrum (\ref{Pr}), 
we obtain the scalar index, its running, and the running of the running in the $\alpha$-vacuum.  
Defining the dimensionless ratio
\begin{equation}
\lambda \equiv \frac{\Lambda}{H_*} \;,
\end{equation}
with $H_*$ evaluated at the pivot scale, we obtain the compact expressions
\begin{eqnarray}
n_s &=& 1 - 6\epsilon_V + 2\eta_V
+ \Delta n_s(\lambda)\;, 
\label{nsalpha}
\\
\alpha_s &=& 16\eta_V\epsilon_V -24\epsilon_V^2 - 2\xi_V^2
+ \Delta\alpha_s(\lambda)\;, 
\label{dnsalpha}
\\
\beta_s &=& 192\eta_V\epsilon_V^2 - 32\eta_V^2\epsilon_V -24\xi_V^2\epsilon_V -192\epsilon_V^3
+ 2\eta_V\xi_V^2 + 2\omega_V^3
+ \Delta\beta_s(\lambda)\;,
\label{d2nsalpha}
\end{eqnarray}
where the $\alpha$-vacuum correction terms are given by
\begin{eqnarray}
\Delta n_s(\lambda) \!\!&=&\!\!
\frac{2\epsilon_V\!\left[(2\lambda^2 - 1)\cos(2\lambda) - 2\lambda\sin(2\lambda)+1\right]}
{-2\lambda^2 + 2\lambda\sin(2\lambda) + \cos(2\lambda) - 1}, \\
\Delta\alpha_s(\lambda) \!\!&=&\!\!
\frac{4\epsilon_V\!\left[(1-2\lambda^2)\cos(2\lambda)+2\lambda\sin(2\lambda)-1\right](2\epsilon_V - \eta_V)}
{2\lambda^2 -2\lambda\sin(2\lambda) - \cos(2\lambda) + 1}, \\
\Delta\beta_s(\lambda) \!\!&=&\!\!
\frac{4\epsilon_V\!\left[(1-2\lambda^2)\cos(2\lambda)+2\lambda\sin(2\lambda)-1\right]
(2\eta_V^2+\xi_V^2 -14\eta_V\epsilon_V +16\epsilon_V^2)}
{2\lambda^2 -2\lambda\sin(2\lambda) - \cos(2\lambda) + 1}.
\end{eqnarray}

Equations~(\ref{nsalpha})–(\ref{d2nsalpha}) generalize the results previously obtained in 
Refs.~\cite{Broy:2016zik,BouzariNezhad:2018zsi} to the full $\alpha$-vacuum-corrected spectrum in Equation~(\ref{Pr}).  
As expected, in the limit $\lambda \to \infty$ (i.e., $\Lambda \gg H_*$), the $\alpha$-vacuum corrections vanish and 
Equations~(\ref{nsalpha})–(\ref{d2nsalpha}) reduce to the standard Bunch--Davies expressions~\cite{Planck:2013jfk}.  
However, for finite $\lambda$, the deviations can be significant, particularly in the spectral 
index $n_s$, potentially leading to observable imprints within current experimental precision.

%%%%%%%%%%%%%%%%%%%%%%%%%%%%%%%%%%%%%%%%%%%%%%%%%%%%%%%%%%%%%%%
\subsection{Tensor Spectrum and the Tensor-to-Scalar Ratio in the $\alpha$-Vacuum}

The presence of a $\alpha$-vacuum also modifies the tensor perturbations, 
since the derivation of the tensor power spectrum parallels that of the scalar sector, 
differing only in the effective mass term and the absence of slow-roll suppression.  
In the BD vacuum, the tensor spectrum at horizon crossing is given by
\begin{equation}
\mathcal{P}_{T}^{\rm BD}(k) = \frac{2H^2}{\pi^2 M_{\rm Pl}^2} \;,
\label{PTBD}
\end{equation}
and the corresponding tensor-to-scalar ratio is~\cite{Planck:2013jfk}
\begin{equation}
r_{\rm BD} = \frac{\mathcal{P}_{T}^{\rm BD}}{\mathcal{P}_{\mathcal{R}}^{\rm BD}}
\simeq 16 \epsilon_V\;,
\label{rBD}
\end{equation}
where the last equality holds at leading order in slow roll.

In the presence of $\alpha$-vacuum modifications, the tensor mode functions undergo 
the same Bogoliubov transformation as in Equation~(\ref{bogo}), with the same coefficients 
$A_k$ and $B_k$ and therefore the same dependence on the ultraviolet (UV) cutoff $\Lambda$.  
Therefore, the corrected tensor spectrum takes the form
\begin{equation}
\mathcal{P}_{T}(k)
= \mathcal{P}_{T}^{\rm BD}(k)
\left[ 1 + e^{\alpha+\alpha^*} + e^{\alpha} + e^{\alpha^*} \right] \;,
\label{PTalpha-generic}
\end{equation}
analogous to the scalar expression.  
Using Equation~(\ref{expalpha}) and the same manipulations leading to Equation~(\ref{Pr}), the tensor spectrum can be written explicitly as
\begin{equation}
\mathcal{P}_{T}(k)
= \frac{2H^2}{\pi^2 M_{\rm Pl}^2}
\left[
1 + \frac{H^2}{2\Lambda^2}
- \frac{H}{\Lambda}
\sin\!\left(\frac{2\Lambda}{H}\right)
- \frac{H^2}{2\Lambda^2}
\cos\!\left(\frac{2\Lambda}{H}\right)
\right] \;,
\label{PTalpha}
\end{equation}
which mirrors the structure of Equation~(\ref{Pr}) for the scalar sector, 
but without the additional slow-roll suppression factors.
Interestingly, because the scalar and tensor power spectra acquire 
identical multiplicative corrections from the $\alpha$-vacuum, these 
factors cancel exactly in the ratio. Note
 that this cancellation 
holds at leading order in slow roll and for scale-independent Bogoliubov 
coefficients; more general initial states or higher-order corrections may break it. 
We, therefore, still obtain the standard slow-roll expression
given in Equation~(\ref{rBD}),
independently of the scale $\Lambda$ that parameterizes the $\alpha$-vacuum modification.
This result has two important implications:
\begin{itemize}
\item[(i)]  
The $\alpha$-vacuum corrections can significantly affect the observables of the 
scalar sector, such as $n_s$, $\alpha_s$, and $\beta_s$, potentially producing 
oscillatory signatures, but they do not alter the prediction of leading order for $r$.

\item[(ii)]
Although both scalar and tensor power spectra acquire identical
multiplicative corrections in the $\alpha$-vacuum, the tensor spectral
index $n_T$ does not remain unchanged.  
Because $n_T$ is defined through the logarithmic derivative of the tensor
spectrum, it receives an additional $\alpha$-dependent contribution~\cite{Broy:2016zik},
while the tensor-to-scalar ratio $r$ does not.  
As a consequence, the standard single-field consistency relation,
\begin{equation}
n_T = -\frac{r}{8} \;,
\label{nT}
\end{equation}
is generally violated at the level of observable quantities.  
Thus, an observationally measurable deviation from $n_T = -r/8$ would
provide a direct signature of non-Bunch--Davies initial conditions.

\end{itemize}

We emphasize, however, that subleading slow-roll corrections or additional 
model-dependent features (e.g., in models with non-canonical kinetic terms 
or modified dispersion relations) may break this exact cancellation.  
In the present framework, the tensor-to-scalar ratio remains unmodified despite potentially 
large oscillatory corrections in the scalar sector. Since the possible violation for the 
consistency relation Equation~(\ref{nT}) due to the quantum vacuum choice is of the same 
order as expected for the spectral index for the scalar perturbations, we will focus 
on the changes in $n_s$ and its runnings.

%%%%%%%%%%%%%%%%%%%%%%%%%%%%%%%%%%%%%%%%%%%%%%%%%%%%%%%%%%%%%%%%%
\subsection{Alpha-Vacuum Corrections at Sub-Planckian Energy Scales}

From the expressions obtained in the previous subsection, it is clear that
the deviation from the BD predictions in the
$\alpha$-vacuum is governed entirely by the dimensionless ratio
$\lambda \equiv \Lambda/H_{*}$.  
In the limit $\lambda \gg 1$, the oscillatory correction terms appearing in the
scalar spectral index and its runnings become highly rapid and average to zero,
thus recovering the standard BD expressions.  
However, for moderate values of $\lambda$, the oscillatory terms remain finite
and produce non-negligible corrections to the observable quantities.  
Therefore, determining whether a tension exists between the BD and $\alpha$
vacua reduces to establishing whether physics permits a value of
$\Lambda/H_{*}$ that is not parametrically large.

Typically, one assumes $\Lambda \gg H$ and identifies the UV cutoff with the
Planck scale, $\Lambda_{\rm Pl}\sim \mathcal{O}(10^{18})\,\mathrm{GeV}$
\cite{Chen:2024ckx,Danielsson:2002kx}.  
Under this choice, the arguments of the sinusoidal functions become enormous,
and the oscillatory terms average out, yielding no observable deviation from
the BD predictions.  
Consequently, a measurable distinction between the two vacua is only possible
if there is a physical mechanism that allows the momentum cutoff to lie at a
significantly lower scale---for example, at the inflationary Hubble scale 
$\Lambda \sim H_{\rm infl} \sim \mathcal{O}(10^{13})\,\mathrm{GeV}$.  
In this regime, $\lambda = \Lambda/H_{*}$ is of order unity, and the
$\alpha$-vacuum corrections yield finite contributions to $n_{s}$,
$\alpha_{s}$, and $\beta_{s}$, producing, in principle, observable deviations.

We therefore conclude that the phenomenological relevance of the
$\alpha$-vacuum hinges upon whether a consistent theoretical framework permits
a sub-Planckian UV cutoff.  
In what follows, we show that the framework of large extra dimensions provides a
natural and well-motivated mechanism for lowering the effective gravitational
scale, thus allowing $\Lambda$ to be as small as the Hubble scale during
inflation.  
This possibility, originally motivated by the hierarchy problem
\cite{Arkani-Hamed:1999wga}, enables the oscillatory corrections to remain
observable and ultimately allows one to distinguish between the BD and
$\alpha$ vacua.

We emphasize that the $\alpha$-vacuum does not resolve the
trans-Planckian problem itself but instead provides an effective
description of possible UV-sensitive corrections, whose observational
imprints can be systematically constrained.

%%%%%%%%%%%%%%%%%%%%%%%%%%%%%%%%%%%%%%%%%%%%%%%%
\subsection{Large Extra Dimensions}
\label{sec3.2.1}

The introduction of large extra spatial dimensions into cosmology is strongly
motivated by attempts to address the hierarchy problem and to explore the
phenomenology of low-scale quantum gravity.  
A considerable body of theoretical work has demonstrated that the presence of
extra compactified dimensions can substantially impact primordial density
perturbations, baryogenesis, reheating, and early-universe dynamics
\cite{Anchordoqui:2023etp,Mohapatra:2000cm,Green:2002wk}.  
To our knowledge, the present work provides the first application of large
extra dimensions to the study of the scalar spectral index and its runnings in
the context of the $\alpha$-vacuum.

Unlike the Planck-sized compact dimensions that arise in typical
Kaluza--Klein or string-theoretic setups, the Arkani-Hamed--Dimopoulos--Dvali
(ADD) scenario allows the compactification scale to be many orders of magnitude
larger than $L_{\rm Pl} \sim 10^{-35}\,\mathrm{m}$
\cite{Arkani-Hamed:1999wga,Witten:1995ex}.  
In a $(3+1+n)$-dimensional bulk spacetime with $n$ compact extra dimensions of
common radius $R$, the Einstein--Hilbert action takes the form
\begin{equation}
    S_{4+n}
    = \int d^4x\, d^ny
    \left\{
        \frac{1}{2} M_{*}^{2+n} \mathcal{R}^{(4+n)}
        - \Lambda_{4+n}
    \right\} \;,
\end{equation}
where $M_{*}$ is the fundamental gravitational scale in the bulk, and
$\mathcal{R}^{(4+n)}$ is the $(4+n)$-dimensional Ricci scalar.  
Here, $\Lambda_{4+n}$ is the higher-dimensional analog of the cosmological
constant; it should not be confused with the momentum-space cutoff appearing
in the $\alpha$-vacuum.

Compactification of the $n$ extra dimensions leads to the standard relation
between the $(4+n)$-dimensional scale $M_{*}$ and the effective four-dimensional
Planck mass,
\begin{equation}
    M_{\rm Pl}^{2}
    \sim M_{*}^{2+n} R^{n} \;,
    \label{eq:Mpl_relation}
\end{equation}
which follows from the higher-dimensional generalization of Gauss's law
\cite{SatheeshKumar:2006it}.  
The simplest compactification topology is $R^{4}\times M_{n}$, where
$M_{n}$ is an $n$-dimensional compact manifold of characteristic volume
$R^{n}$ \cite{Arkani-Hamed:1999wga}.

The primary phenomenological constraint on the size of the extra dimensions
arises from precision tests of Newtonian gravity.  
Cavendish-type experiments constrain deviations from the inverse-square law to
length scales below approximately $250\,\upmu\mathrm{m}$
\cite{Hoyle:2004cw}.  
Using the conversion
\begin{equation}
    1\,\mathrm{mm} \simeq 5.07\times 10^{12}\,\mathrm{GeV}^{-1} \;,
\end{equation}
this yields an upper bound
\begin{equation}
    R < r_{\rm cav}
    \simeq 1.27\times 10^{12}\,\mathrm{GeV}^{-1} \;.
\end{equation}

Rewriting Equation~\eqref{eq:Mpl_relation} as
\begin{equation}
    R
    \sim \frac{1}{M_{*}}
      \left( \frac{M_{\rm Pl}}{M_{*}} \right)^{2/n}
    < r_{\rm cav} \;,
    \label{eq:R_constraint}
\end{equation}
we find the constraint
\begin{equation}
    \left( \frac{M_{\rm Pl}}{M_{*}} \right)^{2/n}
    < r_{\rm cav} M_{*} \;.
    \label{eq:constraint_general}
\end{equation}

\begin{figure}[H]
  %  \centering
    \includegraphics[width=0.5\linewidth]{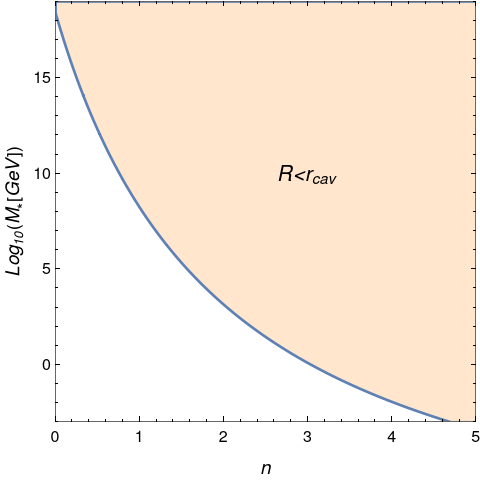}
    \caption{Range of values that satisfy $R < r_{cav}$ in terms of the number of 
    extra spatial dimensions $n$ and the energy scale of gravity on the bulk $M_{*}$.}
    \label{fig1}
\end{figure}

The range of allowed energy scales that satisfy $R < r_{cav}$ is shown on the logarithmic plot in {}Figure~\ref{fig1}.

{}For our purposes, the most relevant value of the bulk scale is the Hubble
scale during inflation, $M_{*} \sim \Lambda_{H} \sim
\mathcal{O}(10^{13})\,\mathrm{GeV}$, corresponding to $\lambda=\Lambda/H\sim
\mathcal{O}(1)$.  
Substituting $M_{*}\sim 10^{13}\,\mathrm{GeV}$ into
Equation~\eqref{eq:constraint_general}, we obtain
\begin{equation}
    \left( 2.44\times 10^{5} \right)^{2/n}
    < 1.27\times 10^{25} \;,
\end{equation}
which implies
\begin{equation}
    n
    > \frac{2\log(2.44\times10^{5})}
           {\log(1.27\times10^{25})}
    \approx 0.43 \;.
    \label{eq:constraint_n}
\end{equation}

Since $n$ must be a positive integer, the condition is trivially satisfied for
all $n\in\mathbb{Z}^{+}$.  
Thus, within the ADD framework, a gravitational scale as low as
$M_{*}\sim H_{\rm infl}$ is entirely consistent with the current experimental
bounds.  
This provides consistent and phenomenologically allowed justification for adopting
a momentum cutoff at the Hubble scale in the $\alpha$-vacuum formalism,
thereby allowing the oscillatory corrections to the primordial spectra to be
observable.

{}Finally, solving Equation~\eqref{eq:R_constraint} for $M_{*}$ yields
\begin{equation}
    \frac{M_{\rm Pl}^{2/n}}{r_{\rm cav}}
    < M_{*}^{(2+n)/n} \;,
\end{equation}
or, equivalently,
\begin{equation}
    \frac{2.44\times 10^{36/n}}{1.27\times10^{12}}
    < M_{*}^{(2+n)/n} \;,
    \label{eq:Mstar_bound}
\end{equation}
demonstrating explicitly the lower bound on the gravitational scale for any
chosen number of extra dimensions $n$.  
This completes the range of allowed values of $M_{*}$ consistent with
experimental bounds and confirms that a cutoff at the inflationary Hubble
scale falls well within the permissible region.
%%%%%%%%%%%%%%%%%%%%%%%%%%%%%%%%%%%%%%%%%%%%%%
\section{Numerical Comparison of Cosmological Observables}
\label{section4}

We now compute explicit predictions for the scalar spectral index $n_s$, its running
$\alpha_s$, and the running of the running $\beta_s$ in the $\alpha$-vacuum using
Equations~(\ref{nsalpha})--(\ref{d2nsalpha}). These predictions are
compared with the corresponding BD values and with the Planck 2018
constraints~\cite{Planck:2018jri}.

\begin{figure}[H]
%\centering
\includegraphics[width=7.3cm]{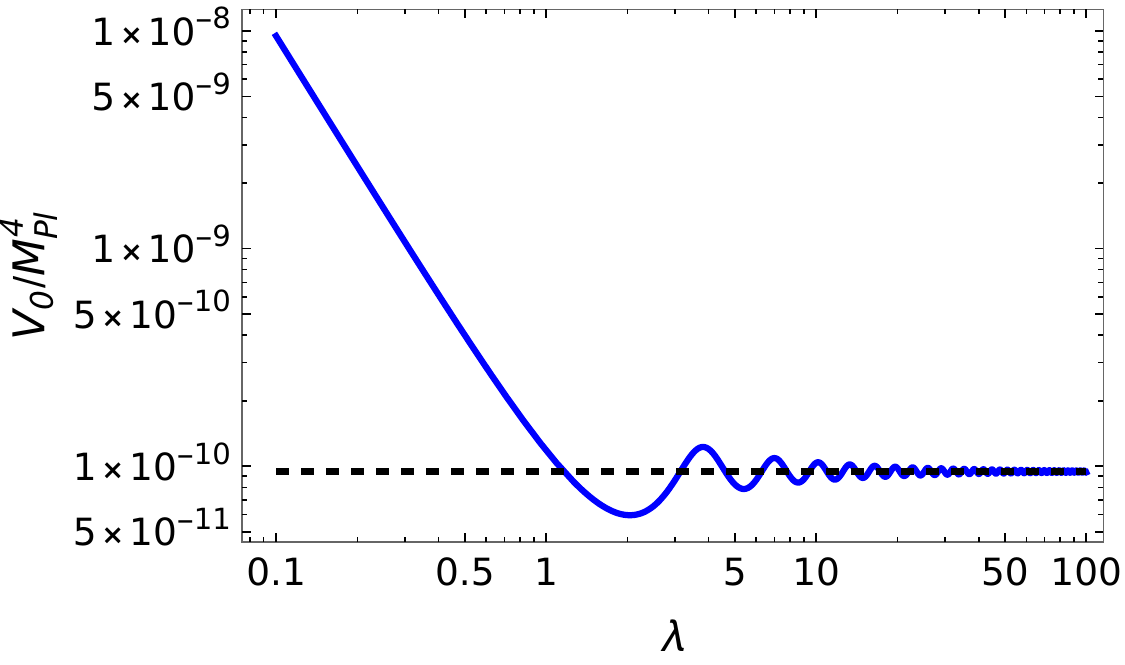}
\caption{
Normalization of the Starobinsky potential as a function of $\lambda$,
computed using the full power spectrum result Equation~(\ref{Pr}). The dashed line
shows the BD limit.}
\label{fig2}
\end{figure}

As a representative inflationary model, we adopt the Starobinsky potential, whose
excellent agreement with current CMB observations makes it an appropriate benchmark
for comparing vacuum choices:
\begin{equation}\label{pot}
    V(\phi)=V_0\!\left(1-e^{-\sqrt{2/3}\,\phi/M_{\rm Pl}}\right)^2,
\end{equation}
where the normalization scale $V_0$ is fixed through the CMB scalar amplitude
$A_s \simeq 2.105\times10^{-9}$ at the pivot scale $k_* = 0.05\,{\rm Mpc}^{-1}$
with a fiducial number of e-folds $N_*=60$. In the BD vacuum, this yields
$V_0^{\rm BD}/M_{\rm Pl}^4 = 9.456\times10^{-11}$. Because the scalar power spectrum
depends on the quantum vacuum, $V_0$ acquires a corresponding $\lambda$-dependence in
the $\alpha$-vacuum, as illustrated in Figure~\ref{fig2}.

%%%%%%%%%%%%%%%%%%%%%%%%%%%%%%%%%%%%%%%%%%
%%%%%%%%%%%%%%%%%%%%%%%%%%%%%%%%%%%%%%%%%%

\begin{figure}[H]
\hspace{-70pt}
\centering
\subfloat[\centering]{\includegraphics[width=5.7cm]{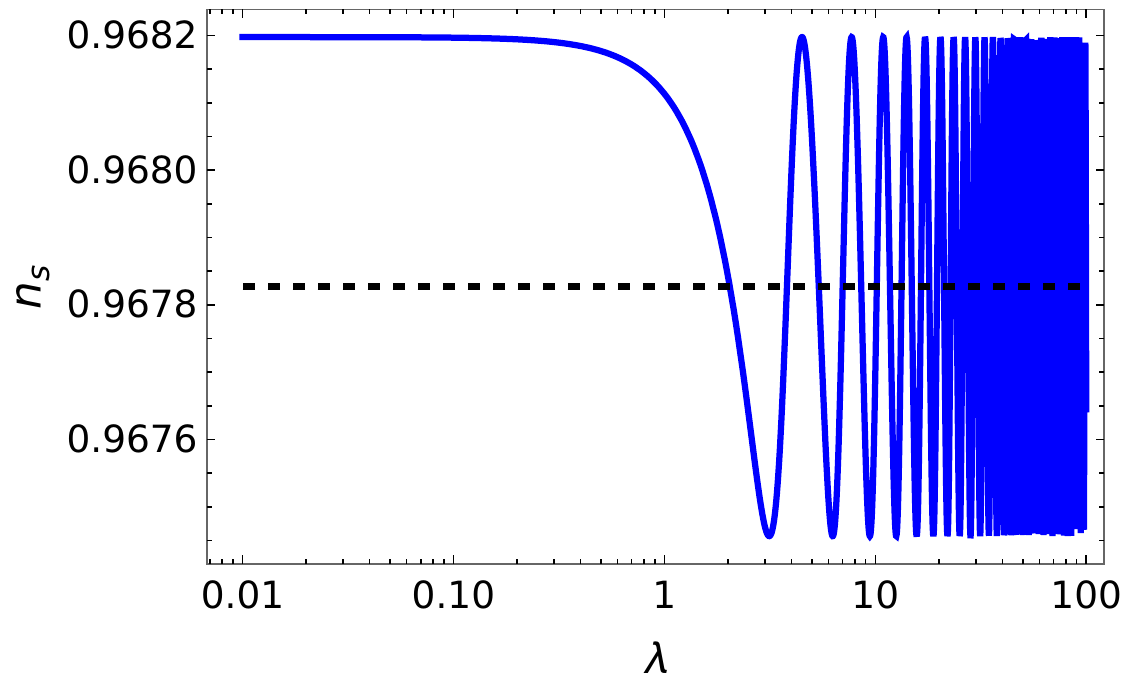}}
\subfloat[\centering]{\includegraphics[width=6cm]{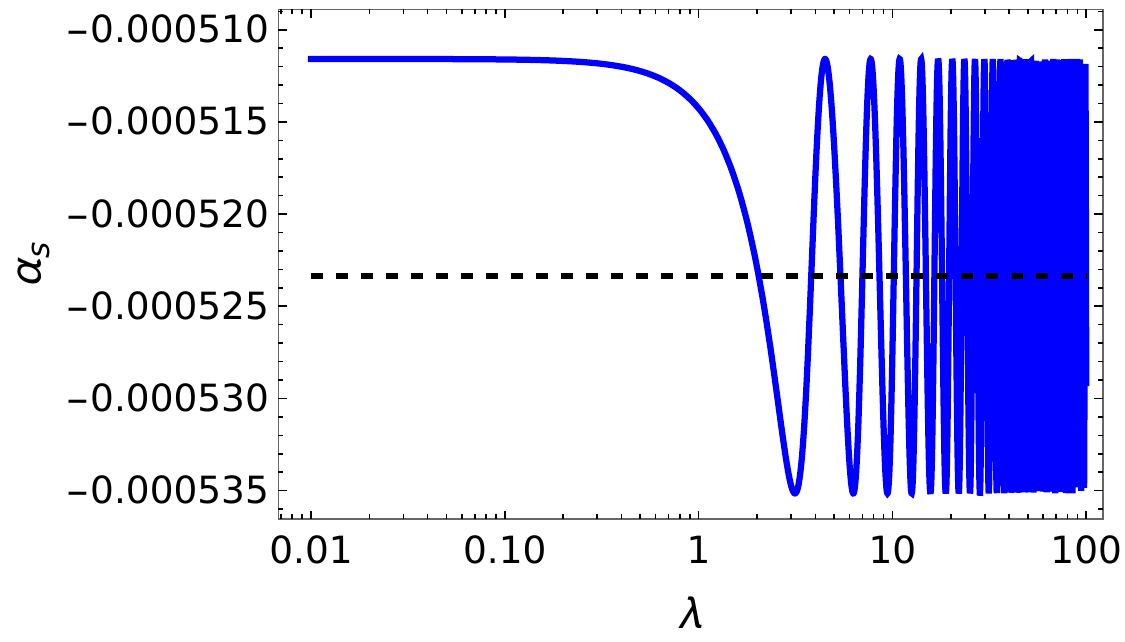}}\\
\hspace{-70pt}
\subfloat[\centering]{\includegraphics[width=6cm]{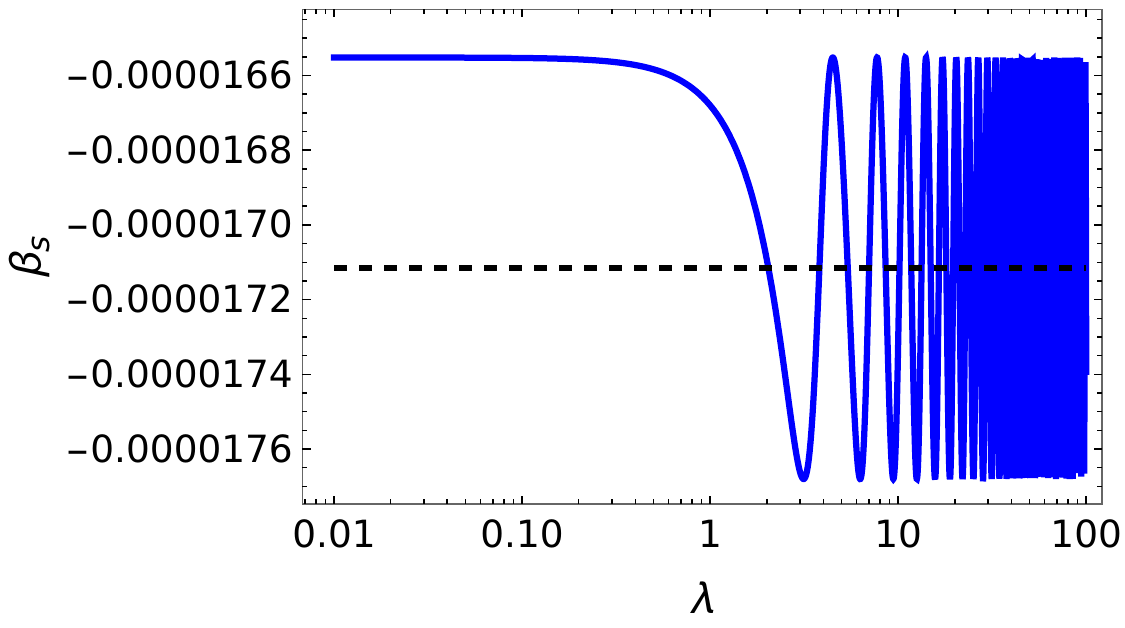}}
\caption{
Predictions for (a) $n_s$, (b) $\alpha_s$, and (c) $\beta_s$ using
Equations~(\ref{nsalpha})--(\ref{d2nsalpha}),
with $N_*=60$. Solid: $\alpha$-vacuum with $\lambda$-dependent corrections.
Dashed: BD limit.}
\label{fig3}
\end{figure}

{}Figure~\ref{fig3} displays the $\lambda$-dependence of $n_s$, $\alpha_s$, and
$\beta_s$ in the Starobinsky model. In each case, the corrections saturate for
$\lambda \lesssim {\cal O}(1)$, corresponding to a cutoff scale near the Hubble
scale during inflation. To quantify the deviations from the BD case, we expand
Equations~(\ref{nsalpha})--(\ref{d2nsalpha}) for small~$\lambda$. Subtracting the BD
expressions yields
\begin{align}
\Delta n_s &= 2\epsilon_V + {\cal O}(\lambda^2), \\
\Delta \alpha_s &= 4\epsilon_V\left(2\epsilon_V - \eta_V\right)
                  + {\cal O}(\lambda^2), \\
\Delta \beta_s &= 4\epsilon_V\!\left(2\eta_V^2 + \xi_V^2
      - 14\eta_V\epsilon_V + 16\epsilon_V^2\right)
      + {\cal O}(\lambda^2) \;,
\end{align}
which provides a good approximation to the maximum corrections extracted
numerically at $\lambda \lesssim 1$.\vspace{-6pt}

%%%%%%%%%%%%%%%%%%%%%%%%%%%%%%%%%%%%%%%%%%
%%%%%%%%%%%%%%%%%%%%%%%%%%%%%%%%%%%%%%%%%%

Table~\ref{table1} summarizes the maximum deviations of the three observables
for $\lambda \lesssim 1$, compared with the BD values and Planck 2018 constraints.
The $\alpha$-vacuum results assume $\Lambda \sim H$ (i.e.,\ $\lambda=\mathcal{O}(1)$),
consistent with the lower-energy cutoff motivated by the large-extra-dimension
mechanism discussed previously. Note that this
 choice sets $\Lambda/H \sim {\cal O}(1)$,
while maintaining consistency with the constraint~(\ref{eq:Mstar_bound}). No preferred
value of the truncation scale is assumed beyond this physical bound.

\begin{table}[H]
%\begin{center}
\caption{\label{table1}
Comparison
 of cosmological observables for the Starobinsky model
between the BD vacuum, the $\alpha$-vacuum (evaluated at its maximal
correction for $\lambda \lesssim 1$), and Planck 2018~\cite{Planck:2018jri,Planck:2018vyg}.
 All results
assume $N_*=60$.}
\setlength{\tabcolsep}{7.2mm}
\begin{tabular}{cccc}
\toprule
\textbf{Parameter} & \textbf{Planck 2018} & \textbf{BD Vacuum} & \textbf{\boldmath{$\alpha$}-Vacuum} \\ \midrule
$n_s$       & $0.9647\pm0.0043$ & $0.9678$ & $0.9682$ \\ \midrule
$\alpha_s$  & $0.0011\pm0.0099$ & $-5.23\times10^{-4}$ & $-5.11\times10^{-4}$ \\ \midrule
$\beta_s$   & $0.009\pm0.012$   & $-1.71\times10^{-5}$ & $-1.66\times10^{-5}$ \\ \bottomrule
\end{tabular}
%\end{center}
\end{table}

The differences between the BD and $\alpha$-vacuum predictions are numerically small,
as expected from the structure of the oscillatory corrections once the cutoff is
constrained to lie near the Hubble scale. Nevertheless, the corrections are finite
and non-zero—reflecting the fact that the $\alpha$-vacuum is observationally
distinguishable but only under the highly restricted circumstances that permit a
sub-Planckian cutoff.

Finally, it is instructive to compare this with the recent ACT results~\cite{ACT:2025tim,ACT:2025fju},
which, when combined with Planck~\cite{Planck:2018jri,Planck:2018vyg} and DESI
data~\cite{DESI:2024mwx,DESI:2024uvr}, give the updated constraint
\[
n_s = 0.974 \pm 0.003 \;.
\]

Even though the $\alpha$-vacuum corrections shift $n_s$ upward relative to the BD case,
the magnitude of the shift for $\Lambda\sim H$ remains too small to reach the new
preferred central value. Thus, while the $\alpha$-vacuum yields consistent predictions
within current uncertainties, it does not significantly alleviate the emerging tension
in~$n_s$ as far as the recent ACT results are concerned.

Although we here have focused on the Starobinsky model as a representative 
and well-motivated example, the structure of the corrections induced by the 
modified vacuum is generic and expected to persist in other slow-roll inflationary models. 
The quantitative impact depends on the background evolution, but the qualitative behavior of the deviations remains robust.

%%%%%%%%%%%%%%%%%%%%%%%%%%%%%%%%%%%%%%%%%%
\section{Conclusions}
\label{conclusions}

In this work, we have examined the impact of replacing the standard BD vacuum with 
the more general $\alpha$-vacuum on the inflationary observables derived from the 
curvature power spectrum. As explicitly shown in Equations~(\ref{nsalpha})--(\ref{d2nsalpha}), the scalar spectral index, its running, and the running of the 
running acquire oscillatory UV-sensitive corrections. These corrections vanish smoothly 
in the limit of large momenta truncation and therefore reproduce the BD expressions when $\Lambda \rightarrow \infty$.

In the usual high-energy effective field theory context, one adopts a cutoff 
at the Planck scale, $\Lambda \sim \mathcal{O}(10^{18})$GeV, which effectively 
eliminates any distinction between the BD and $\alpha$-vacua. Motivated by scenarios 
with large extra dimensions~\cite{Arkani-Hamed:1999wga,Anchordoqui:2023etp,Mohapatra:2000cm,Green:2002wk}, 
we consider instead the possibility of a physical truncation near the Hubble 
scale during inflation, $\Lambda \sim H_{\rm inf} \sim \mathcal{O}(10^{13})$GeV. 
Within this framework, the trans-Planckian corrections are not necessarily Planck-suppressed, and the oscillatory terms appearing in the $\alpha$-vacuum can 
contribute finite, albeit small, modifications to the inflationary observables.

Our detailed numerical analysis shows that deviations from the BD predictions 
remain very small—typically well within the current observational bounds from the 
Planck 2018 data~\cite{Planck:2018jri}—even when the $\alpha$-vacuum corrections 
take their largest possible amplitude. In this sense, the BD vacuum remains a robust 
attractor from the perspective of current CMB measurements. Nevertheless, 
the $\alpha$-vacuum provides a controlled and theoretically consistent way 
to parameterize trans-Planckian effects, offering a phenomenological framework 
to assess potential signatures of non-standard initial states in future high-precision observations.

The consequences of adopting the $\alpha$-vacuum may extend beyond the scalar 
power spectrum. In particular, any observable directly derived from 
$\mathcal{P}_{\mathcal{R}}(k)$ would inherit analogous corrections, 
although their quantitative impact requires dedicated analyses beyond 
the scope of the present work~\cite{Raccanelli:2023zkj,Bolliet:2023sst}.

{}Finally, it is worth noting that the role of vacuum choice is a recurring theme 
in several areas of quantum field theory in curved spacetimes, including black hole physics and horizon thermodynamics~\cite{Hawking:1975vcx,Page:2004xp}. Although the present work 
is restricted to inflationary perturbations, it would be interesting to explore 
whether similar effective parameterizations of initial states could be employed 
in other contexts. We leave such investigations, as well as possible connections to 
holographic frameworks~\cite{Chamblin:2006xd}, for future work.

\vspace{6pt} 

\authorcontributions{Conceptualization, M.W.-S., R.O.R. and A.B.; Methodology, R.O.R. and A.B.; Formal analysis, M.W.-S., R.O.R. and A.B.; Investigation, M.W.-S. and A.B.; Writing---original draft, M.W.-S., R.O.R. and A.B.; Writing---review \& editing, R.O.R. and A.B.; Supervision, R.O.R. and A.B. All authors have read and agreed to the published version of the manuscript.}

\funding{
R.O.R. 
 acknowledges the financial support via research grants from Conselho
Nacional de Desenvolvimento Cient\'{\i}fico e Tecnol\'ogico (CNPq),
Grant No. 307286/2021-5, and from Funda\c{c}\~ao Carlos Chagas Filho
de Amparo \`a Pesquisa do Estado do Rio de Janeiro (FAPERJ), Grant
No. E-26/201.150/2021. 
A.B. is partially funded by STFC.
}

\informedconsent{Not applicable}%MDPI: Any research article describing a study involving humans should contain this statement. Please add ``Informed consent was obtained from all subjects involved in the study.'' OR ``Patient consent was waived due to REASON (please provide a detailed justification).'' OR ``Not applicable'' for studies not involving humans. You might also choose to exclude this statement if the study did not involve humans. Written informed consent for publication must be obtained from participating patients who can be identified (including by the patients themselves). Please state ``Written informed consent has been obtained from the patient(s) to publish this paper'' if applicable.

\dataavailability{The raw data supporting the conclusions of this article will be made available by the authors on request.} 

\conflictsofinterest{The authors declare no conflicts of interest.} 

\abbreviations{Abbreviations}{
~~~~~~~~~~The following abbreviations are used in this manuscript: Bunch--Davies (BD), cosmic microwave background (CMB), Atacama Cosmology Telescope (ACT), confidence level (CL),  ultraviolet (UV), Arkani-Hamed–Dimopoulos–Dvali (ADD).
}

%%%%%%%%%%%%%%%%%%%%%%%%%%%%%%%%%%%%%%%%%%
\begin{adjustwidth}{-\extralength}{0cm}

\reftitle{References}

\PublishersNote{}
\end{adjustwidth}
\end{document}